\documentclass[aps,eqsecnum,nofootinbib,showpacs]{revtex4-1}
\usepackage[dvips]{graphicx}  
\usepackage{amsmath,amsfonts}
\usepackage{bm}
\newcommand{\romanNum}[1]{\@roman{#1}}
\usepackage{color}  
\def\be{\begin{equation}}
\def\ee{\end{equation}}
\def\bea{\begin{equation}\begin{aligned}}
\def\eea{\end{aligned}\end{equation}}
\def\be{\begin{eqnarray}}
\def\ee{\end{eqnarray}}
\def\beq{\begin{equation}}
\def\eeq{\end{equation}}

\def\p{\partial}

\def\({\left (}
\def\){\right )}
\def\[{\left [}
\def\[{\right ]}

\newcommand{\IP}[2]{\big\langle\, #1\, \big\vert\, #2\, \big\rangle}
\newcommand{\Exp}[1]{\langle\, #1\, \rangle}
\bmdefine{\bmk}{\bm{k}} \bmdefine{\bmx}{\bm{x}}
\bmdefine{\bmA}{\bm{A}} \bmdefine{\bmB}{\bm{B}}
\bmdefine{\bmJ}{\bm{J}}

\newcommand{\cO}{\mathcal{O}}
\newcommand{\calL}{\mathcal{L}}
\newcommand{\calO}{\mathcal{O}}
\newcommand{\odiff}[2]{ \frac{d #1}{d #2} }
\newcommand{\odiffII}[2]{ \frac{d^2 #1}{d #2^2} }
\begin{document}

\author{Hua-Bi Zeng$^{1}$, Yu Jiang$^{1}$, Zhe-Yong Fan$^{1}$, and Hong-Shi Zong$^{1,2}$}
\address{$^{1}$ Department of Physics, Nanjing University, Nanjing 210093, China}
\address{$^{2}$ Joint Center for Particle, Nuclear Physics and Cosmology, Nanjing 210093, China}
\title{Characteristic Length of a Holographic Superconductor with $d$-wave Gap}

\begin{abstract}
After the discovery of the $s$-wave and $p$-wave holographic superconductors, holographic models of $d$-wave superconductor have also been constructed recently. We study analytically the perturbation of the dual gravity theory to calculate the
superconducting coherence length $\xi$ of the $d$-wave holographic superconductor near the superconducting phase transition point. The superconducting coherence length $\xi$  divergents as $(1-T/T_c)^{-1/2}$ near the critical temperature $T_c$. We also
obtain the magnetic penetration depth $\lambda\propto(T_c-T)^{-1/2}$ by adding a small external homogeneous magnetic field. The results agree with the $s$-wave and $p$-wave models, which are also the same as the Ginzburg-Landau theory.
\end{abstract}
\pacs{11.25.Tq, 74.20.-z}
\maketitle

\section{ introduction}
The application of AdS/CFT correspondence \cite{1,2,3,4} to condensed matter physics has made much progress in the past two years. Holographic theories of many condensed matter phenomena have been founded, e.g., holographic models of superfluid \cite{5,6,7,8,9,10,11,12,13}, superconductor \cite{14,15,16,17,18,19,20,21,22,23}, quantum phase transition \cite{24,25,26,27,28,29}, fractional quantum hall effect \cite{30,31,32} and topological insulator/superconductor \cite{33}.

The first AdS/CFT model of superconductor is the $s$-wave holographic superconductor, in which a black hole is coupled with a scalar field and Maxwell field \cite{14,15,16}. The conductivity of the superconductor at zero temperature was studied in Ref. \cite{17}. The holographic Fermi surface of the superconductor has been studied by calculating the fermionic spectral function \cite{34,35,36}. Then a $p$-wave
holographic superconductor was also found, in which it is a black
hole coupled with pure $SU(2)$ fields. It is an Einstein-Yang-Mills
theory \cite{18,19}. The non-abelian holographic superconductor away from the
probe limit was studied in Refs. \cite{20,21}. The zero temperature limit of the model
was studied in \cite{22}. The behaviour of fermionic spectral function has also been studied in the p-wave superconducting background \cite{37}. Very recently, two $d$-wave holographic superconductors has been constructed. In these models, the condensed scalar field of the
$s$-wave model is replaced by a symmetric, traceless second-rank tensor in order to capture
the character of the anisotropic $d$-wave order parameter \cite{56,57,23}. This new
holographic superconductor is very interesting since the HTSC is believed to be $d$-wave pairing. Actually, the model reproduces some of
notable properties of cuprates \cite{56,57}.

The behaviors of the $s$-wave holographic superconductor in the presence of
magnetic fields have been studied in many papers \cite{39,40,41,42,43,44,45,46}. In our recent paper \cite{47}, we studied the phase transition properties of
the Einstein-Yang-Mills model in a constant external magnetic field. We found that the
magnetic field added indeed suppresses the superconductivity. Following closely Meada and Okamura \cite{48}, we also study the $p$-wave holographic superconductor by using perturbation theory near the critical temperature. We find that the correlation length $\xi$ diverges as $\xi\sim(1-T_c/T)^{-1/2}$ at the critical temperature. We also find that the magnetic penetration length behaves as $\lambda\sim(T_c-T)^{-1/2}$ near the critical temperature. Both of these results are consistent with the Ginzburg-Landau theory. They
are also the same as the case for the $s$-wave holographic superconductor
which has been studied by Maeda {\it et al} \cite{41}.

In the present paper, in order to see whether the new holographic superconductor model built by Benini, Herzog,  Rahman, and Yaromin in \cite{56,57} with $d$-wave gap also has the mean-field theory critical exponents like the other two models, we study the holographic superconductor's correlated length and magnetic penetration length near the critical phase transition point by the perturbation method which has been used for the $s$-wave and $p$-wave models. We get the expected results which are similar to those of
mean-field theory.

The organization of this paper is as follows. In Section II, we
reconstruct the superconducting solution of the classic gravity
theory which is dual to a $d$-wave superconductor by perturbation
techniques. Section III is devoted to the derivation of $\xi$ by
solving the eigenvalue equations from the perturbation. In Section
IV we find that the London current can be induced by a homogeneous
magnetic field, and the magnetic penetration length is also studied.
The conclusion and some discussions are given in Section V.

\section{Model of a $d$-wave holographic superconductor}
In this section, we will review the gravity dual theory of the
$d$-wave superconductor in \cite{56,57}. In order to have a $d$-wave condensate at the
boundary, we can introduce a charged tensor field with higher spin coupled with a U(1) gauge field in the dual gravity theory.
This is the phenomenological bottom-up construction of holographic superconductor
assuming the existence of gauge/gravity duality.

It is important to note that for higher spin fields, besides the equations of motion, we need additional constraints to remove the unphysical degrees of freedom \cite{48,49,50}. Furthermore, it appears that the construction of higher spin field
coupled to gravity or U(1) gauge field in a gauge invariant way is non-trivial in flat space or curved space \cite{51}.
In spite of this there are some attempts to finish this task \cite{52,53}.
Until now there is no fully consistent theory of interacting higher spin
gauge field.
The difficulties of building a consistent spin two fields coupled to a Maxwell field in the curved background include
the appearance of ghost and causality \cite{54,55,65}. Due to this fact, in \cite{23} the authors proposed a truncated model which has sufficient
ingredients to catch some features of $d$-wave superconductor, leaving the construction of a complete theory to the future.
We would like to emphasize that, in \cite{56,57} Benini, Herzog, Rahman, and Yarom proposed
a similar model of charged spin 2 field under the background of an AdS black hole, in which they find that those inconsistencies
mentioned above can be made very small by taking some limit \cite{57}. This is an important step to achieve a perfect $d$-wave holographic superconductor. In this paper we take their model to be our starting point.

The Lagrangian in 3+1
dimensional curved spacetime which is dual to a 2+1 dimensional $d$-wave
superconductor studied in \cite{56,57} is

\begin{equation}
\begin{split}
\label{E:ActionSimp}
\mathcal{L} &= - |D_\rho \psi_{\mu\nu}|^2 + 2|D_\mu \psi^{\mu\nu}|^2 + |D_\mu \psi|^2 - \big[ D_\mu \psi^{*\mu\nu} D_\nu \psi + \text{c.c.} \big] - m^2 \big( |\psi_{\mu\nu}|^2 - |\psi|^2 \big) \\
&\quad +2  R_{\mu\nu\rho\lambda} \psi^{*\mu\rho} \psi^{\nu\lambda}
- R_{\mu\nu} \psi^{*\mu\lambda} \psi^\nu_\lambda
- \frac{1}{4} R | \psi |^2
- i  q F_{\mu\nu} \psi^{*\mu\lambda} \psi^\nu_\lambda - \frac14 F_{\mu\nu} F^{\mu\nu} \;,
\end{split}
\end{equation}
where $D_\mu = \nabla_\mu - i q A_\mu$ and $\psi_\rho = D^\mu \psi_{\mu\rho}$, $\psi_{\mu \nu}$
is a symmetry traceless tensor.

The equations of motion which follow from (\ref{E:ActionSimp}) are
\bea
\label{EOM probe action}
& 0 = (\square - m^2) \psi_{\mu\nu} - 2 D_{(\mu} \psi_{\nu)} + D_{(\mu} D_{\nu)} \psi - g_{\mu\nu} \big[ (\square - m^2) \psi - D^\rho \psi_\rho \big] \\
&\qquad + 2 R_{\mu\rho\nu\lambda} \psi^{\rho\lambda} - g_{\mu\nu} \frac {R}{4} \psi - i\frac q2 \big( F_{\mu\rho} \psi^\rho_\nu + F_{\nu\rho} \psi^\rho_\mu \big) \\
& D_\mu F^{\mu\nu} = J^\nu
\eea
in which
\be
J^\nu = i \psi^*_{\alpha\beta} (D^\nu \psi^{\alpha\beta} - D^\alpha \psi^{\nu\beta}) + i(\psi^*_\alpha - D_\alpha \psi^*)(\psi^{\nu\alpha} - g^{\nu\alpha} \psi) + \text{h.c.} \;.
\ee

This model gives the correct degrees of freedom for a spin two field\cite{57}.
The Lagrangian \eqref{E:ActionSimp} also has some good properties. For a neutral, non-interacting spin two field in flat spacetime, it reduces to the Fierz-Pauli Lagrangian \cite{48}.  For a neutral spin two field in a fixed Einstein background, ${\mathcal L}$ reduces to that of  Buchbinder- Gitman-
Pershin \cite{65}.
For a charged spin two field in flat spacetime, ${\mathcal L}$ reduces to the Lagrangian of Federbush \cite{66}.
It is also ghost-free, and for generic values of $F_{\mu\nu}$, it describes $5$ propagating degrees of freedom \cite{57}.
Furthermore, this Lagrangian is unique since if we choose the coupling to be different from those in  \eqref{E:ActionSimp}, then some of the
would-be constraints become dynamic, introducing ghosts \cite{57}. However, the equations of motion derived from \eqref{E:ActionSimp} for generic
value of $F_{\mu \nu}$ are either non-hyperbolic or non-causal propagating \cite{55}, as pointed in \cite{57}, the problem can be corrected by
adding to the Lagrangian terms that are high order in $F_{\mu\nu}$. At last, in order to have the right propagating degree of freedom, we must work in
a fixed background spacetime satisfying the Einstein condition \cite{57}
\begin{equation}
\label{EinsteinCond}
R_{\mu\nu} = \frac{2 \Lambda}{d-1} \, g_{\mu\nu}  \;,
\end{equation}
in which $d=3$, the negative cosmological constant $\Lambda=-3$.
This also means that we have to work in the
probe limit where the tensor field and Maxwell field do not backreact the metric.

By taking the planar Schwarzchild-AdS ansatz, the black hole metric satisfying Einstein equation  \eqref{EinsteinCond} reads (we use mostly plus signature for the metric)
\begin{equation}
ds^2 = \frac{L^2}{z^2} \Big(-f(z)\,dt^2 + dx^2 + dy^2 + \frac{dz^2}{f(z)} \Big)
\end{equation}
where
\begin{equation}
f(z) = 1 - \Big( \frac{z}{z_h} \Big)^3 \;.
\end{equation}
The black hole horizon is located at $z=z_h$, while the conformal boundary of the spacetime is located
at $z=0$.
The Hawking temperature of this black hole is
\begin{equation}
T = \frac{3}{4 \pi z_h} \;.
\end{equation}
which is also the temperature of the dual gauge theory living on the
boundary of the AdS spacetime.
Since at zero temperature the back reaction will produce a qualitative difference on the geometry, the approximation of neglecting back reaction of tensor fields is not reliable in the
zero temperature limit \cite{57}.

For the $d$-wave backgrounds, the ansatz takes the following form \cite{56,57}
\begin{equation}
\label{ansatz}
A = A_\mu \, dx^\mu \equiv  \phi(z) \, dt  \;, \qquad\qquad
\psi_{xy}(z) \equiv \frac{L^2}{2z^2} \, \psi(z) \;,
\end{equation}
with all other components of $\psi_{\mu\nu}$ set to zero, and $\phi$ and $\psi$ are real. The ansatz \eqref{ansatz} satisfies $\psi = \psi_\mu = F_{\mu\rho} \psi^\rho_\nu = 0$. Instead of turning on $\psi_{xy}$ in \eqref{ansatz} we could have considered a non-vanishing value for $\psi_{xx-yy} \equiv \psi_{xx} = - \psi_{yy}$. These two ansatzs are equivalent under a
$\pi/4$ rotation \cite{57}.

With this ansatz, we can derive the equations of motion for the two
fields $\psi$ and $\phi$,

\begin{equation}
 \left( z^2 \odiff{}{z}\, \frac{f(z)}{z^2}\, \odiff{}{z}
  - \frac{L^2\, m^2}{z^2} \right)\, \psi
  + \frac{q\phi^2}{f(z)}\, \psi~  =0,
\label{eom1}
\end{equation}
and
\begin{equation}
  f(z)\, \odiffII{\phi}{z}
  - \frac{\, \psi^2}{z^2}\, \phi~ = 0.
 \label{eom2}
\end{equation}

The exact solution of the EOMs which corresponds to the normal state is clearly
\begin{equation}
\psi=0, \phi=\mu(1-\frac{z}{z_h}),
\end{equation}
where $\mu$ is interpreted as the chemical potential of the field theory. The superconducting state with $\psi \neq 0$ exists when $T/q\mu$ is small enough. If we consider the chemical potential as fixed, then the critical value of $T/q\mu$ defines a critical temperature $T_c$ below which the spin two field condenses. For convenience we can also simply set $q = 1$ below.

The superconducting solution with non-vanishing $\psi$ takes
the following asymptotically form at the AdS boundary \cite{56,57}:
\begin{equation}
\psi(z) =  z^{d-\Delta} \big[ \psi^{(s)}  + O(z) \big] + z^{\Delta} \Big[ \frac{\langle \cO_{xy} \rangle}{2 \Delta - 3}   + O(z) \Big] \;,
\end{equation}

\begin{equation}
\phi = \mu- \rho z + \cdots~,
\end{equation}
where $\rho$ is the charge density of the boundary field theory, $\Delta$ is given by $m^2 L^2 = \Delta(\Delta - 3)$, $m^2\geq0$\cite{57}.
We can check that such choice of dimension is compatible with unitarity bounds in conformal theories, for instance \cite{68}.
Also in \cite{57}, the authors give a generalization of the analysis of Breitenlohner and Freedman \cite{67} to obtain the bound of $m^2\geq0$.

We can take $\psi^{(s)}$ as the source, then $\langle \cO_{xy} \rangle$ can be taken as the vacuum expectation value (VEV) of the operator that couples to the
tensor field in the boundary theory, which is also the order parameter of the superconducting phase.
We require that $\psi^{(s)}$ equal zero so that the spin two field is not sourced.

According to numerical calculations for the case $\Delta=4$ (in this case $m^2 L^2=4$) in Ref. \cite{57}, the order parameter behaves as
\begin{equation}
\langle \cO_{xy} \rangle \sim(1-T/T_c)^{1/2}
\end{equation}
near the critical phase transition point. For reasons of
continuity, the solution at the critical temperature should be
\begin{equation}
\psi_c=0, \phi_c=\mu_c(1-\frac{z}{z_h}).
\end{equation}
The non-trivial solution near the critical temperature can be obtained by
a perturbation expansion in terms of $\epsilon=(1-T/T_c)$ since
$\epsilon$ is a small parameter. We expand $\psi(z)$ and
$\phi(z)$ as
\begin{equation}
   \psi(z)
  = \epsilon^{1/2}\,\psi_1(z)
  + \epsilon^{3/2}\, \psi_2(z) + \cdots~,
  \label{exp1}
\end{equation}
\begin{align}
 \phi(z)
  = \phi_c(z) + \epsilon\, \phi_1(z) + \cdots~.
\label{exp2}
\end{align}

Substituting Eq. (\ref{exp1}) and Eq. (\ref{exp2}) into
Eq. (\ref{eom1}) and Eq. (\ref{eom2}), we obtain equations for
$\psi_1$ and $\phi_1$:
\begin{equation}
{\calL}_\psi \psi_1(z)=0, \label{eq.expand1}
\end{equation}

\begin{equation}
\frac{d^2 \phi_1(z)}{dz^2}-\frac{\psi_1^2 \phi_c}{z^2f(z)}=0. \label{eq.expand2}
\end{equation}

Here, for convenience, in the next two sections we define:
\begin{equation}
{\cal
L}_\psi=-\left( z^2\frac{d}{dz}\frac{f(z)}{z^2}\frac{d}{dz}-\frac{L^2 m^2}{z^2}+\frac{\phi_c^2}{f(z)} \right).
\end{equation}

\section{The Superconducting Coherence Length }
The superconducting coherence length is obtained by the complex pole of the static
correlation function of the order parameter in Fourier space:
\begin{align}
  & \Exp{\Tilde{\calO}(\Vec{k}\,) \Tilde{\calO}(-\Vec{k}\,)}
  \sim \frac{1}{ \Vec{k}^2 + 1/\xi^2}~.
\end{align}
The pole $ \Vec{k}^2$ can be calculated in the probe
limit by perturbing the fields ($\psi$, $\phi$) in the
equations of motion . It is
enough to consider a linear perturbation with fluctuation of the
fields in the $x$-direction which takes the following form:
\begin{equation}
\delta \phi(z,x)dt=[\Phi(z,k)dt]e^{ikx},
\end{equation}
\begin{equation}
\delta \psi(z,x)=[\Psi(z,k)]e^{ikx}.
\end{equation}

Using this perturbation, we get the following linearized equations
for $\Psi$ and $\Phi$:
\begin{equation}
k^2\Psi=(z^2\frac{d}{dz}\frac{f(z)}{z^2}\frac{d}{dz}-\frac{L^2 m^2}{z^2}+\frac{\phi^2}{f(z)})\Psi+\frac{2\psi \phi}{f(z)}\Phi,
\end{equation}
\begin{equation}
k^2\Phi=(f(z)\frac{d^2}{dz^2}-\frac{\psi^2}{z^2})\Phi-\frac{2\phi\psi}{z^2} \Psi,
\end{equation}

Now, our task is to solve the eigenvalue equations near $T_c$
analytically. Using the perturbation expansions in Eq. (\ref{exp1})
and Eq. (\ref{exp2}), we get
\begin{equation}
-k^2\Psi=({\calL}_\psi- \frac{2 \epsilon \phi_c \phi_1}{f})\Psi-\frac{2\epsilon^{1/2}\phi_c \psi_1}{f(z)}\Phi,
\label{eq.eigein1}
\end{equation}
\begin{equation}
-k^2\Phi=(-f(z)\frac{d^2}{dz^2}+\frac{\epsilon \psi_1^2}{z^2})\Phi+\frac{2\epsilon^{1/2} \phi_c\psi_1}{z^2}\Psi.
\label{eq.eigein2}
\end{equation}
The boundary conditions for the two equations are: at the horizon,
\begin{equation}
\Psi(1)=\textrm{regular}, ~~~~~\Phi(1)=0,
\end{equation}
near the AdS boundary $z=0$,
\begin{equation}
\Psi(z)=(\textrm{const})\times z^4+O(z^8)~,
\end{equation}
\begin{equation}
\Phi(z)=(\textrm{const})\times z+O(z^2)~.
\end{equation}

The trivial solution is the zeroth order solution $\Phi_0$ and $\Psi_0$
with $k=0$:
\begin{equation}
\Phi_0=0, ~~~~\Psi_0=\psi_1,
\end{equation}
in which we have used the Eq. (\ref{eq.expand1}). The non-trivial solutions can be obtained by a series expansion around the zeroth order solution in $\epsilon$:
\begin{equation}
\Psi=\psi_1+\epsilon \Psi_1+\epsilon^2 \Psi_2+\cdots
\end{equation}
\begin{equation}
\Phi=\epsilon^{1/2}\Phi_1+\epsilon^{3/2}\Phi_2+\cdots
\end{equation}
\begin{equation}
 k^2=\epsilon k^2_1+\epsilon^2 k^2_2+\cdots
\end{equation}
Using this expansion in Eq. (\ref{eq.eigein1}) and Eq. (\ref{eq.eigein2}), one has
\begin{equation}
-k^2_1
\psi_1={\cal L}_\psi \Psi_1-\frac{2\phi_c\psi_1}{f(z)}(\phi_1+\Phi_1),
\label{eq.k}
\end{equation}

\begin{equation}
\frac{d^2\Phi_1}{dz^2}=\frac{2\phi_c\psi_1^2}{z^2 f(z)}=\frac{2d^2\phi_1}{dz^2},
\end{equation}
where the last equality on the right comes 
directly from Eq.(II.19).

Eq. (\ref{eq.k}) can be solved for $k$ by defining an inner
product for the state $\psi_I$ and $\psi_{II}$:
\begin{equation}
 \IP{\psi_I}{\psi_{II}}= \int_0^{z_h} \frac{dz}{z^2}\psi_I^*(z)~\psi_{II}(z).
\end{equation}

Using this inner product for Eq. (\ref{eq.k}) and $\psi_1$,
together with the fact that ${\cal L}_\psi \psi_1=0$, we have

\begin{equation}
k^2_1\IP{\psi_1}{\psi_1}=\left\langle
\psi_{1}|\frac{2 \phi_{c}\psi_{1}}{f(z)}
\phi_{1}\right\rangle +2\int_{0}^{z_h}dz\frac{\phi_{c}\psi_{1}^{2}}{z^2 f(z)}\Phi_{1}.
\end{equation}

The first term of the above equation vanishes, which can be seen
from the Hermiticity of ${\cal L}_\psi$ and
\begin{equation}
{\cal L}_\psi \psi_{2}=\frac{2
\phi_{c}\psi_{1}}{f(z)} \phi_{1}. \label{eq.w2}
\end{equation}
Eq. (\ref{eq.w2}) is the equation of motion for $\psi_2$ defined
in Eq. (\ref{exp1}). Using the fact that $ k^{2}=\epsilon  k_{1}^{2}$, in a  first order approximation the eigenvalue $k$  may be written as
\begin{equation}
  k^{2}=\epsilon\frac{N}{D} + O(\epsilon^{2}),
\end{equation}
 where
\begin{eqnarray}
 N=2\int_{0}^{z_h}dz\frac{ \phi_{c}\psi_{1}^{2}}{z^2 f(z)}\Phi_{1}=-\frac{1}{2}\int_{0}^{z_h}dz(\frac{d \Phi_1}{dz})^2 < 0,
   \\~~~\textrm{ and} ~~~D=\int_{0}^{z_h}dz\frac{\psi_{1}^{2}}{z^2} > 0.
\end{eqnarray}

For the last equality in Eq.(III.21), we used Eq.(III.16) and boundary
condition Eq.(III.9) and Eq.(III.10). Finally, the superconducting coherence length is given by
\begin{equation}
\xi = \frac{\epsilon^{-1/2}}{\beta(T_c)}\sqrt{-\frac{D}{N}}
+O(\epsilon^{2})\propto \left( 1 - \frac{T}{T_c} \right)^{-1/2}~.
\end{equation}
We have thus obtained the same critical exponent $-1/2$ for $\xi$ as
in Ginzburg-Landau theory. In this sector, the calculations are focused on the case $\Delta=4$, which
corresponds to $m^2L^2=4$. In fact, the change of mass will not change the Hermiticity of $\cal L_{\psi}$, and therefore will not
change the result we obtained. We can extend the calculations to the general mass case for $m^2>0$ and the
same critical exponents are expected.

\section{The London Equation And The Magnetic Penetration Depth}
To calculate the magnetic penetration length for the holographic
superconductor, we add a small homogenous external magnetic field by
assuming a perturbative non-vanishing vector potential $\delta
A_{y}(z,x)=b(z)x$, where $ \lim_{z \to 0}\, \delta
A_{y}(z,x)=Bx$. Then the magnetic field in the field theory is
$F_{xy} = \p_x \delta A_{y}= B$ \cite{40,47}. We still work in the
probe limit where the magnetic field does not affect the metric. If
we only focus on the neighborhood of $x=0$, the equation of motion
for $b(z)$ can be treated as decoupled from $\psi$:
\begin{equation}
(\frac{d}{dz}f(z)\frac{d}{dz}-\frac{\psi^2(z)}{z^2})b(z)=0. \label{eq.b}
\end{equation}
$b(z)$ must satisfies the regularity boundary condition at the
horizon $z=z_h$. This equation can also be solved by perturbation methods. We can
expand $b(z)$ as
\begin{equation}
b(z)=b_0(z)+\epsilon b_1(z)+ \cdots.
\end{equation}
Substituting this expansion and Eq.(\ref{exp1}) into Eq.(\ref{eq.b}),
we obtain the equations
\begin{equation}
\frac{d}{dz}f\frac{d}{dz}b_0(z)=0, \label{eq.b1}
\end{equation}
\begin{equation}
\frac{d}{dz}f\frac{d}{dz}b_1(z)-\frac{\psi_1^2(z)}{z^2}b_0(z)=0.
\label{eq.b2}
\end{equation}
The solution of Eq. (\ref{eq.b1}) which satisfies the required boundary conditions is
\begin{equation}
b_0(z)=C,
\end{equation}
where $C=B$ is a constant since the condition $\lim_{z \to 0}\, b(z)=B$
must be satisfied. So the solution of Eq. (\ref{eq.b2}) should be:

\begin{equation}
\frac{d b_1}{dz}=-\frac{B}{f(z)}\int_z^{z_h}dz_0
\frac{\psi_1^2(z_0)}{z_0^2}.
\end{equation}
Integrating the above equation, we have
\begin{equation}
b(z)=B-\epsilon B\int^z_0 \frac{dz_1}{f(z_1)}
    \int^{z_h}_{z_1} dz_0~\frac{\psi_1^2(z_0)}{z_0^2}
  + O(\epsilon^2).
  \label{eq.b(z)}
\end{equation}
Using the fact that $B = \lim_{z \to 0}\, b(z)$ and $\delta A_y^{(0)}(x) = \lim_{z \to 0}\, \delta A_y(z, x)$, we can rewrite Eq. (\ref{eq.b(z)}) as
\begin{equation}
\delta A_y (z, x) =
          \delta A_y^{(0)}(x)\,\left( 1
  - \, \epsilon\, \int^z_0 \frac{dz_1}{f(z_1)}
    \int^{z_h}_{z_1} dz_0~\frac{\psi_1^2(z_0)}{z_0^2} \right)
  + O(\epsilon^2).
\end{equation}

According to the AdS/CFT dictionary, we can read out the current
$<J_y(x)>$ near $T_c$ to be:
\begin{flalign}
<J_y(x)> \propto -T_c(1-\frac{T}{T_c})\int_0^{z_h}dz \frac{\psi_1^2(z)}{z^2}\delta
A_y^{(0)}(x) + O(\epsilon^2),
\end{flalign}
or
\begin{equation}
<J_y(x)>\sim-T_c\epsilon\delta A_y^{(0)}(x).
\label{eq.J}
\end{equation}
This is similar to the London equation for real world superconductors:
\begin{align}
  & \bmJ = - \frac{e_*^2}{m_*}~\psi^2\, \bmA
  = - e_*\, n_s\, \bmA~,
\label{eq:London_eq}
\end{align}
where $e_*$ and $m_*$ are effective charge and mass of the order
parameter, and $n_s$ is the superfluid number density.

Comparing Eq. (\ref{eq.J}) and Eq. (\ref{eq:London_eq}), we find that the superfluid density $n_s$ near the critical point in the field theory is
\begin{align}
  & n_s \sim \epsilon\, T_c \sim T_c-T~.
\label{magnetic penetration depth}
\end{align}
According to Ginzburg-Landau theory, the magnetic penetration depth
$\lambda$ is given as

\begin{align}
  & \lambda \sim 1/\sqrt{n_s}~.
\end{align}
Then, we get the behavior of $\lambda$ in the vicinity of the
critical temperature
\begin{equation}
\lambda\propto(T_c-T)^{-1/2},
\end{equation}
which is the expected result as in Ginzburg-Landau theory.

\section{ conclusion and discussion}
For a $d$-wave holographic superconductor, the superconducting
coherence length and magnetic penetration depth are studied around
the critical temperature. Their critical exponents both show mean
field behavior, as is expected in advance. The same exponents
appear in both $s$-wave and $p$-wave models though their models are
different. It is believed that the mean-field behavior is due to the fact
that we are using AdS/CFT under the large $N$ limit, in which the gravity theory is
classic and quantum fluctuation is suppressed. Nevertheless, although still in the large $N$ limit, several holographic models show non-mean-field behaviors and the
Berezinskii-Kosterlitz-Thouless phase transition \cite{58,59,60,61,62}. It is an interesting matter to study this issue further to answer when mean-field behavior is expected. The suppression of quantum fluctuation can also explain why the
continuous symmetry breaking can happen in the 2+1 dimensional field theory at a finite temperature, while this is impossible in real word as argued by Coleman-Mermin-Wagner-Hohenberg theorem. Recently, a one-loop correction to the expectation value for the order parameter of the $s$-wave model was studied in Ref.\cite{63}. It is found that the quantum fluctuation of a hydrodynamic second sound mode will indeed wash out the order parameter in the 2+1 dimensional holographic superconductor. It might be interesting to check their general structure ``planar $\textrm{AdS}_4$ horizon + hydrodynamic mode = strong IR fluctuation''
in this $d$-wave model. Furthermore, for the $s$-wave model under a magnetic field, vortex lattice solutions have been found \cite{44}. We will discuss the vortex lattice solution for the new $d$-wave holographic superconductor in the near future \cite{64}.

\section{acknowledgement}

We would thank Wei-Min Sun for valuable comments.
This work is supported in part by the National Natural Science
Foundation of China (under Grant Nos. 10775069, 11075075 and 10935001) and the Research Fund
for the Doctoral Program of Higher Education (Grant No. 20080284020).

\end{document}